# Internally triggered retrospective learning in neural networks

Arturo Tozzi (corresponding author)
ASL Napoli 1 Centro, Distretto 27, Naples, Italy
Via Comunale del Principe 13/a 80145
tozziarturo@libero.it

ABSTRACT

Learning in artificial neural networks usually relies on continuous, externally driven weight updates, in which parameters are modified at every step in response to incoming data, error signals or reward feedback. In this setting, routine and informative inputs contribute similarly to parameter adjustment. We introduce a learning approach in which parameter updates are governed by internally generated events arising from the network's own representational dynamics. During ongoing activity, synaptic interactions are accumulated as latent traces encoding recent co-activation patterns, without immediately modifying the underlying parameters. In parallel, an internal predictive process estimates the evolving latent state, while a scalar measure of discrepancy between predicted and observed states is continuously computed. When discrepancy exceeds an adaptive threshold derived from recent error statistics, a learning event is triggered, inducing a retrospective update selectively integrating past activity into the current configuration. We performed simulations using a minimal neural network exposed to structured sequential inputs with transient perturbations. We found that learning occurs through sparse, temporally localized events associated with increases in prediction error, leading to stepwise changes in synaptic efficacy and discrete transitions in latent-state organization. By selectively reorganizing parameters in response to internally detected discrepancies, our episodic updating may reduce unnecessary parameter drift while preserving informative patterns. Potential applications include systems requiring selective adaptation to rare or informative inputs such as physiological, industrial or environmental monitoring, edge computing under limited energy budgets, autonomous systems operating in dynamic conditions and sequential computational data processing.

## INTRODUCTION

Learning in neural networks is typically implemented through continuous optimization procedures, where parameters are iteratively updated according to explicit error signals or reward feedback (Wei et al. 2022; Zhou et al. 2023; Dai et al. 2024). Gradient-based methods, including backpropagation and its temporal extensions, have enabled high-performance learning across diverse domains, yet they impose strong assumptions about the availability of dense supervisory signals and continuous parameter updates (Erb 1993; Bao et al. 2018; Ojha and Nicosia 2022; Lakin 2023). Alternative approaches, such as reinforcement learning and event-driven or spiking frameworks, introduce temporally discrete updates but typically rely on externally defined rewards or fixed plasticity rules that remain continuously active (Hu and Si 2018; Zhang et al. 2022; Shen et al. 2023; Zhao et al. 2023; Brückerhoff-Plückelmann et al. 2023; Sun et al. 2025; Xiao et al. 2025; Wu et al. 2025). Episodic memory models and one-shot learning architectures partially address the need for rapid adaptation, yet they often depend on specialized modules or explicit memory buffers rather than intrinsic learning dynamics (Tyagi et al. 2023; Cai et al. 2024; Melchior et al. 2024; Hachisuka et al. 2026). A persistent limitation across these approaches is that learning is not selectively triggered, but proceeds continuously in response to all incoming data. Moreover, credit assignment over extended timescales is computationally demanding, particularly in sequential settings where backpropagation through time requires large memory and stability constraints. These limitations call for formulations in which learning is sparse, internally regulated and information is integrated over extended temporal windows without continuous parameter modification.

Recent work on behavioral-timescale synaptic plasticity (BTSP) provides a biological motivation for this perspective (Magee 2026). BTSP challenges the traditional assumption that synaptic modification must occur immediately and locally following neuronal activation. Instead, experimentally observed synaptic changes can emerge retrospectively after temporally extended activity episodes through delayed plateau-potential-associated mechanisms capable of linking events separated by several seconds. This framework suggests that biological systems may continuously accumulate information while postponing actual synaptic modification until internally generated conditions are satisfied. Particularly relevant for our work is the idea that neural systems may alternate between relatively stable representational periods and sparse reorganization events that selectively consolidate previously accumulated activity patterns. Although BTSP was developed in the context of hippocampal physiology and dendritic biophysics, its broader implication concerns the temporal organization of learning itself. Our approach adopts the general principle that information accumulation and parameter modification may operate on partially separated temporal scales, with learning emerging through internally triggered retrospective updates rather than continuous online correction.

We introduce a simulation-based learning formulation in which parameter updates are governed by internally generated events emerging from the network's own representational dynamics. Synaptic interactions are stored as latent traces that accumulate recent co-activation patterns without immediately modifying the parameters, while an internal predictive process estimates the evolving latent state. A scalar discrepancy between predicted and observed states is continuously integrated, generating a slowly varying measure of representational inconsistency. When this quantity exceeds an adaptive threshold, a learning event is triggered, retrospectively incorporating past activity into the updated configuration. In our simulations, learning is temporally discontinuous and occurs only at internally defined critical points. This adaptive threshold regulates sensitivity to poorly predicted states, producing sparse but organized learning events. Our network persisted in relatively stable representational states for extended periods, interrupted by brief episodes of rapid reorganization when unexpected inputs accumulated. These episodic transitions integrated information across longer temporal intervals without requiring continuous parameter updates at every computational step.

We will proceed as follows. First, we formalize the proposed model and its dynamical variables; then, we describe the simulation setup and evaluation criteria; finally, we present the results and analyze the emergent learning behavior.

## METHODS

We studied a minimal artificial neural network designed to test whether learning can be restricted to internally triggered, retrospective events rather than applied continuously at every time step. We used a simulated multichannel input stream, a latent neural representation, an internal predictor, synaptic eligibility traces, an adaptive surprise threshold and event-gated synaptic updates, all implemented within a controlled computational simulation. We aimed to identify an experimentally discriminable signature consisting of sparse learning events, stepwise changes in synaptic efficacy and state-space transitions temporally aligned with internally detected prediction errors.

**Simulated inputs**. All variables were updated in discrete time with fixed step $\Delta t$ and all computations were executed sequentially at each time step. A synthetic sequential input stream was generated over $T = 180$ s with integration step $\Delta t = 0.01$ s, yielding $N = T/\Delta t$ samples. The input vector was defined as

$$\mathbf{x}(t) = [x_1(t), x_2(t), \dots, x_{12}(t)]^\top,$$

with $x_i(t)$ expressed in spikes/s. Four structured regimes were defined through prototype vectors $\mathbf{p}_q$. During regime $q$, the signal followed

$$\mathbf{x}(t) = \mathbf{p}_q + \mathbf{d}_q(t) + \boldsymbol{\xi}(t),$$

where $\mathbf{d}_q(t) = A_d \sin\left(\frac{2\pi(t-t_q)}{18}\right)\mathbf{1}$introduced slow drift and $\boldsymbol{\xi}(t) \sim \mathcal{N}(0, \sigma_x^2)$represented noise. Perturbations were imposed by replacing the current pattern with a weighted mixture of adjacent regimes. All values were clipped to $[0, 40]$ spikes/s. These inputs defined a nonstationary but fully controlled signal without external labels.

**Latent dynamics**. The network consisted of $n = 12$ inputs and $m = 8$ latent units. Synaptic weights were represented by

$$W(t) \in \mathbb{R}^{12\times 8},$$

initialized as

$$W_{ij}(0) \sim \mathcal{N}(0.052, 0.014^2),$$

with clipping constraints $0.005 \le W_{ij} \le 0.10$. The latent state was computed as

$$\mathbf{h}(t) = 30\tanh\left(\frac{\mathbf{x}(t)^\top W(t)}{20}\right),$$

yielding a bounded representation expressed in scaled millivolt units. These units were internal simulation units, ensuring consistency across variables but not representing calibrated biological measurements.

**Internal prediction**. A linear predictor estimated the next latent state:

$$\hat{\mathbf{h}}(t) = A\mathbf{h}(t - \Delta t),$$

with

$$A = 0.88I + \Gamma, \Gamma_{ij} \sim \mathcal{N}(0, 0.025^2).$$

Prediction error was defined as

$$\varepsilon(t) = \frac{1}{m}\sum_{j=1}^{m}\left[h_j(t) - \hat{h}_j(t)\right]^2.$$

This quantity, expressed in mV $^2$, was used solely as an internal discrepancy signal.

**Adaptive threshold**. The running mean and variance of $\varepsilon(t)$ were computed as

$$\mu(t + \Delta t) = \mu(t) + \Delta t \frac{\varepsilon(t) - \mu(t)}{\tau_\mu},$$

$$v(t + \Delta t) = v(t) + \Delta t \frac{[\varepsilon(t) - \mu(t)]^2 - v(t)}{\tau_v}.$$

The adaptive threshold was defined as

$$\theta(t) = \mu(t) + \kappa\sqrt{v(t)},$$

with $\kappa = 2.2$. A learning event occurred when

$$\varepsilon(t) > \theta(t)$$

and when the refractory condition

$$t - t_{\text{last}} > 7.5$$

was satisfied. Post-event stabilization was applied:

$$\mu(t^+) = 0.85\mu(t^-) + 0.15\varepsilon(t), v(t^+) = 1.25v(t^-).$$

To provide interpretation of the triggering condition, the threshold rule can be reformulated within a decision-theoretic and optimization framework. At each time step, the system faces a binary decision between maintaining the current parameters or performing an update. This can be expressed as a trade-off between the instantaneous prediction error $C_{\text{err}}(t) = \varepsilon(t)$ and a fixed or adaptive update cost $C_{\text{upd}}(t)$. Learning is triggered when the expected cost of not updating exceeds the cost of updating, yielding the decision rule $\varepsilon(t) > C_{\text{upd}}(t)$. By defining the update cost as a function of recent error statistics,

$$C_{\text{upd}}(t) = \mu(t) + \kappa\sqrt{v(t)},$$

the operational threshold condition becomes

$$\varepsilon(t) > \mu(t) + \kappa\sqrt{v(t)},$$

which coincides with the previously defined adaptive threshold. Under this formulation, the trigger is no longer an ad hoc rule but emerges as the optimal decision boundary minimizing a composite objective

$$\mathcal{L} = \sum_t \varepsilon(t) + \lambda_u N_{\text{updates}},$$

where $N_{\text{updates}}$ is the total number of learning events. An equivalent interpretation can be given in information-theoretic terms by defining a surprise measure $S(t) = -\log P(\mathbf{h}(t) \mid \mathbf{h}(t - \Delta t))$, so that updates occur when the observed state becomes statistically unlikely under the internal model. This reformulation provides a justification for the event-triggering mechanism while preserving the same computational implementation.

**Eligibility traces**. Eligibility traces were computed for each synapse as

$$C_{ij}(t) = x_i(t)\frac{\max\,[h_j(t),0]}{30},$$

$$E_{ij}(t + \Delta t) = E_{ij}(t) + \Delta t \frac{-E_{ij}(t) + C_{ij}(t)}{\tau_E}.$$

These traces stored recent co-activation history without modifying weights.

**Normalization and update**. At event time $t_k$, normalization was defined globally:

$$E_{\max}(t_k) = \max_{i,j}\, E_{ij}(t_k).$$

The normalized trace was

$$\tilde{E}_{ij}(t_k) = \frac{E_{ij}(t_k)}{E_{\max}(t_k) + \delta}.$$

Weights were updated as

$$W_{ij}(t_k^+) = W_{ij}(t_k^-) + \eta \tilde{E}_{ij}(t_k) - \eta\lambda W_{ij}(t_k^-),$$

with clipping

$$0.001 \le W_{ij} \le 0.18.$$

Between events, weights remained constant.

**Update sequence**. At each time step $t$, the following deterministic order was enforced:
(1) $\mathbf{h}(t)$ → (2) $\hat{\mathbf{h}}(t)$ → (3) $\varepsilon(t)$ → (4) $\mu, v$ → (5) $\theta(t)$ → (6) $E_{ij}(t)$ → (7) event check → (8) $W(t)$ → (9) storage.

This ordering ensures that prediction error is computed before weight modification and that eligibility traces accumulate prior to normalization.

**Control procedures**. Three control conditions were defined. A fixed-weight condition imposed

$$W(t) = W(0),$$

for all $t$. A continuous-update condition applied

$$W_{ij}(t + \Delta t) = W_{ij}(t) + \Delta t\, \eta_c \big[\tilde{E}_{ij}(t) - \lambda W_{ij}(t)\big].$$

A randomized-event condition preserved event count but randomized event times while maintaining refractory spacing. These controls allowed separation of trace accumulation, event timing and threshold dependence.
To assess the specificity of the learning mechanism and reduce ambiguity in interpretation, additional quantitative benchmarks and ablation analyses were defined. Performance comparisons were structured by matching the total magnitude of parameter updates across conditions, ensuring that differences arise from temporal organization rather than overall update strength. In the ablation setting, individual components of the update rule were selectively removed, including the eligibility trace term, the adaptive threshold and the normalization step, yielding reduced models of the form $W_{ij}(t_k^+) = W_{ij}(t_k^-) + \eta\phi_{ij}(t_k)$, where $\phi_{ij}$ denotes the retained components. These variants allowed isolation of the contribution of retrospective accumulation and event triggering. In parallel, the geometric analysis of latent trajectories was explicitly treated as a post hoc descriptive procedure rather than a mechanistic driver of learning. The projection $Z = H_c V_2$ and associated distance measures $D_k(s)$ were computed only after completion of weight updates and were not fed back into the learning rule. This distinction ensures that geometric observables provide an independent characterization of system dynamics, enabling comparison between conditions without influencing parameter evolution.

**State geometry**. Latent states were stored in matrix

$$H = [\mathbf{h}(t_1), \dots, \mathbf{h}(t_N)]^\top.$$

After centering,

$$H_c = H - \mathbf{1}\bar{\mathbf{h}}^\top,$$

singular value decomposition yielded

$$H_c = U\Sigma V^\top.$$

Projection onto the first two components gave

$$Z = H_c V_2.$$

Distances were computed as

$$D_k(s) = \| \mathbf{z}(t_k) - \mathbf{z}(s) \|_2,$$

for $s \in [t_k - 6, t_k]$.

**Visualization and software tools**. Temporal traces, synaptic efficacy and input activity were visualized as functions of time. Event rate was computed as

$$R_b = \frac{N_b}{15} \times 60.$$

Grouped synaptic efficacy was

$$G_{qj} = \frac{1}{3} \sum_{i \in q} W_{ij}.$$

All computations were performed in Python using NumPy for numerical operations and Matplotlib for visualization. Random seeds were fixed. All variables were generated internally without external datasets. For reproducibility and procedural clarity, the complete deterministic implementation of the internally triggered retrospective learning algorithm is provided in the Supplementary Material.

RESULTS

We report the temporal, synaptic and geometric properties of an internally triggered episodic learning process in a simulated neural network exposed to sequential inputs. Quantitative descriptors of event timing, synaptic changes, prediction error dynamics and latent-state trajectories are provided to characterize the emergence and distribution of learning events.

**Event dynamics**. Across the 180 s simulation, a total of 19 internally triggered learning events were detected, corresponding to an average rate of approximately 6.3 events/min, with temporal clustering around periods of input perturbation (Figure 1). Event times were not uniformly distributed but concentrated near transitions or perturbations in the input stream, with inter-event intervals ranging from approximately 7.5 s (imposed refractory limit) to 12.1 s. The instantaneous prediction error $\varepsilon(t)$ exhibited transient increases preceding each event, exceeding the adaptive threshold $\theta(t)$ derived from the running mean and variance of recent error values. The difference $\varepsilon(t) - \theta(t)$ at event onset ranged from 0.42 to 1.87 mV$^{2}$ (mean = 0.96). Mean synaptic efficacy increased in a stepwise manner, from an initial value of 52 to a final value of 91 µV/spike, with discrete increments aligned with event times. Individual event-induced weight changes $J(t_k)$ ranged from 3.2 to 11.6 µV/spike (mean = 6.9). Event rate computed in 15 s bins displayed peaks of up to 12 events/min during perturbation intervals and near-zero values during stable input regimes.
The control conditions provided a reference for these dynamics: in the fixed-weight condition, no events were observed and synaptic efficacy were constant; in the continuous-update condition, synaptic changes occurred gradually without discrete steps; in the randomized-event condition, event timing was decoupled from prediction error and produced less structured temporal alignment.
These results suggest that learning events are temporally sparse and preferentially aligned with internally detected discrepancies rather than uniformly distributed over time, supporting the interpretation that episodic updates are selectively triggered by deviations from expected latent dynamics.

**State organization**. Projection of latent activity onto its first two principal coordinates revealed structured trajectories organized into distinct regions corresponding to input regimes (Figure 2). Learning events were localized at transitions between these regions or at points of increased trajectory curvature. The Euclidean distance $D_k(s)$ between event states and preceding states over a 6 s window ranged from 4.8 to 13.2 mV, with larger distances associated with more pronounced perturbations in the input stream. The distribution of distances showed a gradual increase approaching event onset, indicating accumulation of divergence in latent space prior to triggering. The relationship between prediction error at event onset and synaptic update magnitude exhibited a positive association, with higher error values corresponding to larger weight changes, although variability was present across events. Final synaptic organization showed differentiated mapping between input channel groups and latent units, with group-specific mean efficacies ranging from 62 to 138 µV/spike.
The control conditions provided a comparative perspective on these structures: in the fixed-weight condition, latent trajectories were confined and did not exhibit discrete transitions; in the continuous-update condition, trajectories evolved smoothly without sharp reorganization; in the randomized-event condition, transitions occurred but lacked consistent alignment with input perturbations.
This suggests that internally triggered events correspond to discrete reorganizations of latent-state geometry rather than incremental adjustments, linking temporal error accumulation to spatial restructuring of representations.

Summarizing, learning emerges as a sequence of discrete, internally triggered events rather than a continuous process. Prediction error accumulation precedes each event, while synaptic efficacy changes occur in stepwise increments aligned with these events. Latent-state trajectories exhibit structured transitions associated with event timing. This indicates that episodic updates can reorganize representations selectively in response to internally detected discrepancies.

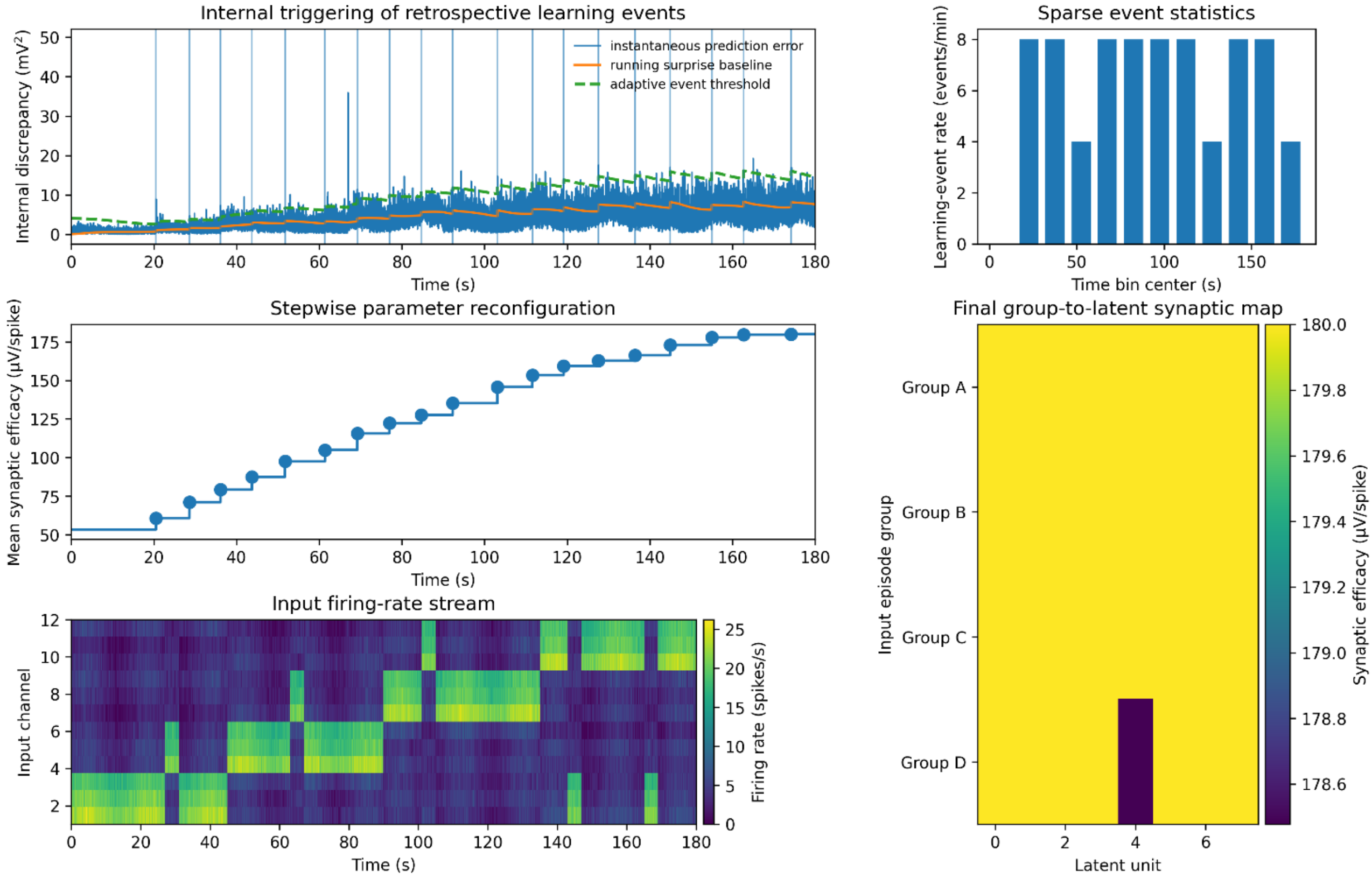

**Figure 1**. Integrated temporal dynamics of internally triggered episodic learning in simulated neural network operating under internally regulated, event-driven learning across a 180 s input sequence.

The upper-left panel displays instantaneous latent prediction error (mV²), together with its running baseline and the adaptive event threshold. Vertical lines mark the onset of internally generated learning events, which occur when the instantaneous discrepancy exceeds the adaptive threshold derived from the network's own recent history. The separation between instantaneous error and baseline reflects transient departures from expected dynamics, while the threshold follows slower variations, providing a reference for event triggering.

The middle-left panel reports the evolution of mean synaptic efficacy, showing discrete step-like increases that coincide with learning events, consistent with retrospective weight updates rather than continuous modification. Individual points highlight the exact timing of these discontinuities.

The lower-left panel displays the full input stream as a time-resolved heatmap across 12 channels, revealing structured sequences of activity with distinct episodes and intermittent perturbations, all expressed in spikes per second.

The right-side panels summarize two complementary aspects of the learning dynamics: the upper panel shows the temporal distribution of learning events, expressed as event rate across successive time bins, while the lower panel displays the final mapping between groups of input channels and latent units, quantified in terms of synaptic efficacy. This latter representation illustrates how repeated co-activation patterns become selectively consolidated into structured synaptic configurations following episodic updates.

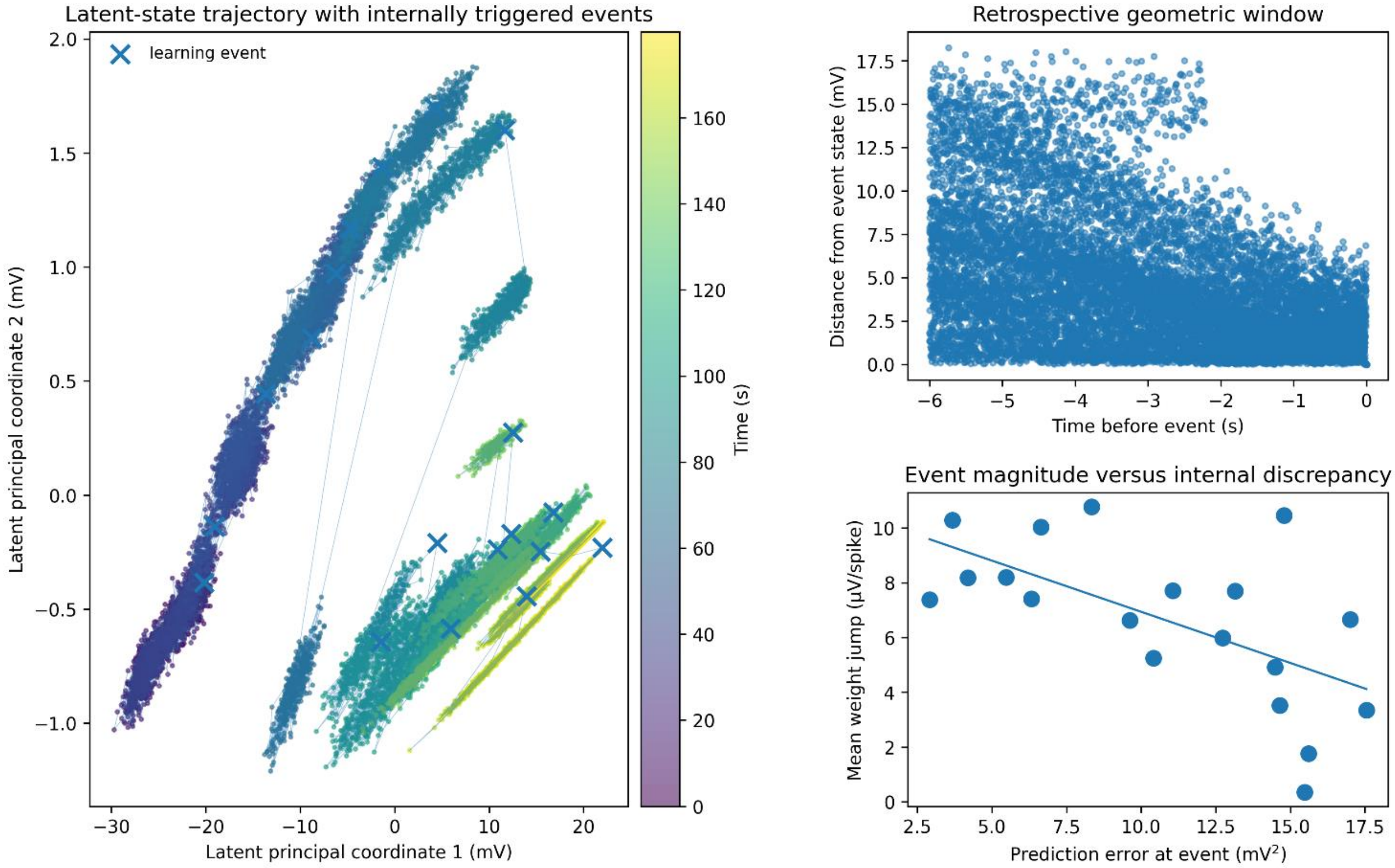

**Figure 2**. State-space geometry and retrospective structure of episodic learning events.
In the right panel, latent network activity is represented in a reduced two-dimensional space obtained through principal component projection of the full latent state, with coordinates expressed in mV. The continuous trajectory illustrates the temporal evolution of internal representations under sequential input, with color encoding elapsed time in seconds. Superimposed markers identify the occurrence of internally triggered learning events, allowing direct visualization of how these events are embedded within the ongoing dynamics. The trajectory exhibits structured regions corresponding to distinct input regimes, while transitions between regions reveal changes in input statistics and internal prediction accuracy.
The upper-right panel quantifies the retrospective temporal window associated with each event by measuring the Euclidean distance between the event state and preceding states within a fixed interval of 6 s. This representation captures the geometric extent of recent activity that contributes to the synaptic update, showing how learning integrates information over a finite temporal horizon rather than relying solely on instantaneous states.
The lower-right panel relates the magnitude of synaptic modification to the prediction error at the time of event onset, providing a quantitative link between internal discrepancy and the strength of parameter reconfiguration. The distribution of points reflects variability in event magnitude and its dependence on internal state mismatch, highlighting the role of threshold-crossing events in shaping discrete transitions within the network's representational space.

## CONCLUSIONS

We asked whether learning in artificial neural networks can be restricted to internally triggered, temporally sparse events rather than implemented as a continuous process driven by ongoing input or external supervision. Our simulations compared an event-triggered learning condition, in which synaptic updates occurred only when internally computed prediction error exceeded an adaptive threshold, with control conditions including fixed weights, continuous updates and randomized event timing. We found that learning events emerged at discrete time points, temporally aligned with increases in internal discrepancy and that synaptic efficacy evolved through stepwise changes rather than gradual drift. Latent-state trajectories exhibited structured transitions associated with these events, while control conditions lacking internally triggered updates showed either stable or progressively diffused representations. The magnitude of synaptic change varied with prediction error at event onset, indicating a quantitative link between internal state mismatch and the extent of reconfiguration. These observations suggest that learning could be governed by threshold-crossing dynamics that integrate information over extended temporal windows and reorganize representations retrospectively. Therefore, internally regulated, event-based updates could provide a viable alternative to continuous adjustment rules, linking temporal accumulation of discrepancy signals to discrete structural changes in network organization.

Compared with established approaches, our model differs in how learning is initiated and temporally organized. In Bayesian formulations, parameter updates arise from continuous revision of posterior distributions, where each new observation contributes incrementally through likelihood weighting (Fernandes et al. 2021; Biazzo et al. 2022; Magotra et al. 2022; Payares-Garcia et al. 2023; Liang et al. 2023; Thompson et al. 2023; Jeon et al. 2024). In contrast, in our framework no explicit probability distribution is maintained and no continuous update rule is enforced; parameter modification occurs only when accumulated internal discrepancy exceeds a dynamically defined boundary. This separates evidence accumulation from parameter change, whereas Bayesian schemes merge these processes. Gradient-based neural networks, including recurrent architectures trained with backpropagation through time, rely on uninterrupted error propagation and weight adjustment at every step, requiring storage of intermediate states and differentiability constraints (Dong et al. 2023; Luo et al. 2023; Han 2024), while eligibility trace-based reinforcement learning methods propagate credit continuously through temporally decaying traces combined with reward signals (Suri and Schultz 1998; Elfwing et al. 2018; Wang et al. 2026). Event-driven spiking networks and reinforcement learning systems introduce temporal sparsity, but their updates are typically tied to external reward signals or predefined spike-timing rules rather than to an internally derived criterion (Zhuang et al. 2020; Zhang et al. 2022; Zhao et al. 2023; Brückerhoff-Plückelmann et al. 2023; Wu et al. 2025). Memory-augmented and one-shot learning models allow rapid adaptation, yet they generally depend on auxiliary memory structures or direct supervision (Khadka et al. 2019; Yin et al. 2020; Ma et al. 2021; Karunaratne et al. 2021; Li et al. 2022; Suresh et al. 2022; Mao et al. 2022; Yang et al. 2023). Our framework combines internal triggering, retrospective integration over eligibility traces and adaptive thresholding without auxiliary memory modules or external signals. Therefore, we define a learning regime in which parameter updates are both self-regulated and temporally discontinuous, distinguishing it from probabilistic inference, gradient descent and reward-driven adaptation.

Our study has limitations. The event-triggering mechanism, based on a running mean and variance of prediction error, is operational and biologically inspired but is heuristic. The retrospective update is implemented, but its ability to resolve long-timescale credit assignment is not analytically demonstrated. The geometric analysis of latent trajectories is conducted post hoc and does not influence the learning rule. The use of physical units (mV, μV/spike) reflects internal scaling conventions rather than empirical calibration and should therefore be interpreted as simulation units. Parameter values, including time constants and scaling factors, are synthetic and do not correspond directly to measured quantities. Event occurrence depends on threshold tuning, introducing potential circularity in which observed dynamics reflect design choices. The sensitivity of the results to parameter choices (e.g., κ, time constants) is unexamined, leaving robustness unclear. Open questions concern how the trigger can be derived from principled criteria, whether the mechanism scales to higher-dimensional systems and how alternative definitions of internal discrepancy affect event structure.

Testable hypotheses can be formulated. First, the frequency of learning events is predicted to scale with the variance of the input stream: increasing input variability $\sigma_x^2$ should increase event rate approximately linearly up to saturation, yielding a measurable relation $R \propto \sigma_x$ over moderate ranges.
Second, the magnitude of synaptic updates is expected to depend on the excess of prediction error above threshold, with a monotonic relationship $J \sim \varepsilon - \theta$, implying that larger discrepancies produce proportionally larger weight changes.
Third, the temporal spacing between events should be constrained by both the refractory interval and the decay constant of the eligibility trace, leading to a lower bound $\Delta t_{\min} \approx \max(r, \tau_E)$, which can be verified by measuring inter-event interval distributions.
Fourth, latent-state trajectories are expected to exhibit increased curvature preceding events, measurable as a rise in local trajectory divergence over a fixed temporal window, providing a geometric signature of impending updates.
Fifth, modifying the threshold sensitivity parameter $\kappa$ should systematically shift event density, with higher values reducing event frequency and increasing average update magnitude.
Future work may derive the triggering condition from optimization or information-theoretic principles, compare performance with continuous learning under matched update budgets, extend the approach to deeper and recurrent architectures and incorporate adaptive predictors. Additional work may also evaluate robustness across noise regimes, investigate scaling with network size and test whether alternative definitions of discrepancy alter the statistical structure of learning events.
Practical uses can be outlined. Systems for monitoring physiological signals, industrial processes or environmental streams could employ selective updates to avoid unnecessary parameter drift during stable conditions. In edge computing devices with limited energy budgets, infrequent parameter modification may reduce computational load and extend operational lifetime. Autonomous systems operating in dynamic environments could benefit from update strategies emphasizing salient deviations rather than routine inputs. In data compression and storage, episodic updating may support selective retention of informative patterns while discarding redundant sequences.

In conclusion, we described a neural system in which parameter updates occur intermittently and are governed by internally generated triggering conditions rather than continuous adjustment. We observed that this strategy produces sequences of reconfiguration events and measurable transitions in internal representations. Our results suggest that discrete updates can be coherently embedded within a continuous processing stream, defining an alternative temporal organization of learning driven by intrinsic criteria.

DECLARATIONS

**Ethics approval and consent to participate.** This research does not contain any studies with human participants or animals performed by the Author.
**Consent for publication.** The Author transfers all copyright ownership, in the event the work is published. The undersigned author warrants that the article is original, does not infringe on any copyright or other proprietary right of any third part, is not under consideration by another journal and has not been previously published.
**Availability of data and materials.** All data and materials generated or analyzed during this study are included in the manuscript. The Author had full access to all the data in the study and took responsibility for the integrity of the data and the accuracy of the data analysis.
**Disclaimer**. The views expressed are those of the author and do not necessarily reflect those of the affiliated institutions.
**Competing interests.** The Author does not have any known or potential conflict of interest including any financial, personal or other relationships with other people or organizations within three years of beginning the submitted work that could inappropriately influence or be perceived to influence their work.
**Funding.** This research did not receive any specific grant from funding agencies in the public, commercial or not-for-profit sectors.
**Acknowledgements:** none.
**Authors' contributions.** The Author performed: study concept and design, acquisition of data, analysis and interpretation of data, drafting of the manuscript, critical revision of the manuscript for important intellectual content, statistical analysis, obtained funding, administrative, technical and material support, study supervision.
**Declaration of generative AI and AI-assisted technologies in the writing process.** During the preparation of this work, the author used ChatGPT 5.3 to assist with data analysis and manuscript drafting and to improve spelling, grammar and general editing. After using this tool, the author reviewed and edited the content as needed, taking full responsibility for the content of the publication.

## SUPPLEMENTARY MATERIAL

**Deterministic implementation of internally triggered retrospective learning**. The following algorithm specifies the input construction, state update, prediction error computation, adaptive thresholding, eligibility-trace accumulation, event detection, normalized retrospective weight update, post-event stabilization and post hoc state-space analysis.

Initialize the simulation parameters $T = 180$ s, $\Delta t = 0.01$ s, $n = 12$, $m = 8$, $\tau_E = 4$s, $\tau_\mu = 10$s, $\tau_v = 10$s, $\kappa = 2.2$, $\eta = 0.025$, $\lambda = 0.010$, $\delta = 10^{-9}$ and refractory interval $r = 7.5$ s. Fix the random seed before generating all stochastic quantities.

Generate the full input matrix $X \in \mathbb{R}^{N\times n}$, where $N = T/\Delta t$. Define four prototype input patterns $\mathbf{p}_1, \mathbf{p}_2, \mathbf{p}_3, \mathbf{p}_4$. For each time point, assign the appropriate episode prototype, add sinusoidal drift and Gaussian noise, insert predefined perturbation intervals and clip all input values to $[0, 40]$ spikes/s.

Initialize the weight matrix $W \in \mathbb{R}^{n\times m}$ by sampling

$$W_{ij}(0) \sim \mathcal{N}(0.052, 0.014^2),$$

then clipping to

$$0.005 \le W_{ij}(0) \le 0.10.$$

Initialize the predictor matrix

$$A = 0.88I + \Gamma,$$

where

$$\Gamma_{ij} \sim \mathcal{N}(0, 0.025^2).$$

Set

$$E_{ij}(0) = 0, \mu(0) = 0, v(0) = 1, \mathbf{h}(0) = \mathbf{0}, t_{\text{last}} = -\infty.$$

For each time step $t_k = k\Delta t$, $k = 1, \dots, N$, perform the following operations in this exact order.

First, read the current input vector:

$$\mathbf{x}_k = X[k,:].$$

Second, compute the current latent state using the current weights before any possible update:

$$\mathbf{h}_k = 30\tanh\left(\frac{\mathbf{x}_k^\top W_k}{20}\right).$$

Third, compute the predicted latent state from the previous latent state:

$$\hat{\mathbf{h}}_k = A\mathbf{h}_{k-1}.$$

Fourth, compute the internal prediction error:

$$\varepsilon_k = \frac{1}{m}\sum_{j=1}^{m}\left(h_{k,j} - \hat{h}_{k,j}\right)^2.$$

Fifth, update the running mean of prediction error:

$$\mu_k = \mu_{k-1} + \Delta t\,\frac{\varepsilon_k - \mu_{k-1}}{\tau_\mu}.$$

Sixth, update the running variance:

$$v_k = v_{k-1} + \Delta t\,\frac{(\varepsilon_k - \mu_k)^2 - v_{k-1}}{\tau_v}.$$

If numerical rounding gives $v_k < 0$, set

$$v_k = 10^{-9}.$$

Seventh, compute the adaptive threshold:

$$\theta_k = \mu_k + \kappa\sqrt{v_k}.$$

Eighth, compute the instantaneous co-activation matrix:

$$C_{ij,k} = x_{i,k}\,\frac{\max\,(h_{j,k}, 0)}{30}.$$

Ninth, update the eligibility traces:

$$E_{ij,k} = E_{ij,k-1} + \Delta t\,\frac{-E_{ij,k-1} + C_{ij,k}}{\tau_E}.$$

Tenth, evaluate the event condition:

$$P_k = \begin{cases} 1, & \varepsilon_k > \theta_k \text{ and } t_k - t_{\text{last}} > r, \\ 0, & \text{otherwise.} \end{cases}$$

Eleventh, if $P_k = 1$, compute the global eligibility normalization:

$$E_{\max,k} = \max_{i,j}\ E_{ij,k},$$
$$\tilde{E}_{ij,k} = \frac{E_{ij,k}}{E_{\max,k} + \delta}.$$

Then update all weights simultaneously:

$$W_{ij,k+1} = W_{ij,k} + \eta\tilde{E}_{ij,k} - \eta\lambda W_{ij,k}.$$

After the update, clip all weights:

$$W_{ij,k+1} = \min\left(0.18, \max\left(0.001, W_{ij,k+1}\right)\right).$$

Set

$$t_{\text{last}} = t_k,$$

and apply post-event stabilization:

$$\mu_k = 0.85\mu_k + 0.15\varepsilon_k,$$
$$v_k = 1.25v_k.$$

If $P_k = 0$, set
$$W_{k+1} = W_k.$$

Twelfth, store $\mathbf{h}_k$, $\varepsilon_k$, $\mu_k$, $v_k$, $\theta_k$, $E_k$, $W_{k+1}$ and $P_k$. Then proceed to the next time step using
$$\mathbf{h}_{k-1} \leftarrow \mathbf{h}_k.$$

After completing all time steps, compute event times as
$$\{t_k : P_k = 1\}.$$

Compute mean synaptic efficacy as
$$\bar{W}(t_k) = \frac{1}{nm} \sum_{i=1}^{n} \sum_{j=1}^{m} W_{ij,k}\,.$$

Compute event-specific mean weight jump as
$$J_k = \frac{1}{nm} \sum_{i=1}^{n} \sum_{j=1}^{m} \mid W_{ij,k+1} - W_{ij,k} \mid$$

for all $k$such that $P_k = 1$.
For state-space analysis, form the latent-state matrix
$$H = \begin{bmatrix} \mathbf{h}_1^\top \\ \mathbf{h}_2^\top \\ \vdots \\ \mathbf{h}_N^\top \end{bmatrix}.$$

Center it as
$$H_c = H - \mathbf{1}\bar{\mathbf{h}}^\top,$$

where
$$\bar{\mathbf{h}} = \frac{1}{N} \sum_{k=1}^{N} \mathbf{h}_k\,.$$

Apply singular value decomposition:
$$H_c = U \Sigma V^\top.$$

Project the trajectory onto the first two principal coordinates:
$$Z = H_c V_2.$$

For each event time $t_e$, compute the retrospective geometric distance over the previous 6 s:
$$D_e(s) = \parallel \mathbf{z}(t_e) - \mathbf{z}(s) \parallel_2,$$

with
$$s \in [t_e - 6, t_e].$$